# Insight into the structural and magnetotransport properties of epitaxial α-Fe$_2$O$_3$/Pt(111) heterostructures – the role of the reversed layer sequence

A. Kozioł-Rachwał[1*], N. Kwiatek[2], W. Skowroński[3], K. Grochot[1,3], J. Kanak[3], E. Madej[2], K. Freindl[2], J. Korecki[2], N. Spiridis[2]

[1]Faculty of Physics and Applied Computer Science, AGH University of Science and Technology, al. Mickiewicza 30, 30-059 Kraków, Poland

[2]Jerzy Haber Institute of Catalysis and Surface Chemistry, Polish Academy of Sciences, 30-239 Kraków, Poland

[3]Institute of Electronics, AGH University of Science and Technology, Al. Mickiewicza 30, 30-059 Kraków, Poland

*e-mail: akoziol@agh.edu.pl

We report on the chemical structure and spin Hall magnetoresistance (SMR) in epitaxial α-Fe$_2$O$_3$(hematite)(0001)/Pt(111) bilayers with hematite thicknesses of 6 nm and 15 nm grown by molecular beam epitaxy on a MgO(111) substrate. Unlike previous studies that involved Pt overlayers on hematite, the present hematite films were grown on a stable Pt buffer layer and displayed structural changes as a function of thickness. These structural differences (the presence of a ferrimagnetic phase in the thinner film) significantly affected the magnetotransport properties of the bilayers. We observed a sign change of the SMR from positive to negative when the thickness of hematite increased from 6 nm to 15 nm. For α-Fe$_2$O$_3$(15 nm)/Pt, we demonstrated room-temperature switching of the Néel order with rectangular, nondecaying switching characteristics. Such structures open the way to extending magnetotransport studies to more complex systems with double asymmetric metal/hematite/Pt interfaces.

## I. INTRODUCTION

Spintronic devices based on antiferromagnets (AFMs) offer a wide range of unique properties, such as robustness against magnetic field perturbations, fast spin dynamics and lack of stray fields. [1] However, the absence of a net magnetic moment in AFM materials is a challenge for electrical reading of their magnetic state, particularly when the AFM layer is insulating. In the last decade, the spin Hall magnetoresistance (SMR) [2] was used to probe the direction of spins in insulating AFMs in contact with a metal characterized by a finite spin Hall angle. The SMR effect, characterized by the dependence of the resistance of the metallic thin film with strong spin-orbit coupling on the direction of the magnetization orientation of the adjacent magnetic layer, was first demonstrated in collinear ferrimagnet/heavy metal (HM) bilayers [3], [4], [5], [6], [7], [8] and applied to ferromagnet (FM)/HM structures. [9] The SMR effect manifests itself as sinusoidal oscillations of the Pt resistance in response to rotation of the ferrimagnet magnetization induced by an external magnetic field. For canted and spiral ferrimagnets ($Gd_3Fe_5O_{12}$ and $Cu_2OSeO_3$, respectively), the SMR was used to identify noncollinear magnetic phases. [10], [11] More recently, the SMR was reported in heterostructures comprising AFMs. Shang *et al.* investigated how the proximity of NiO with different thicknesses influences the SMR signal in the $Pt/NiO/Y_3Fe_5O_{12}$ stack. Calculations performed for the AFM/HM bilayer showed that the SMR can be used to determine the magnetic anisotropy and the Néel order in the AFM layer, [12] for which, despite the absence of net magnetization, a nonzero SMR arises, which is an effect of the individual interaction of magnetic sublattices with the spin current. SMR in the AFM/HM bilayer was experimentally shown for $Ta/Cr_2O_3$ [13] and Pt/NiO. [14], [15], [16] For AFMs, a 90° phase shift (so-called "negative" SMR) was observed compared to the SMR observed in ferrimagnet/Pt bilayers (known as "positive" SMR) due to the orthogonal alignment of the AFM spins with respect to the external magnetic field exceeding the spin-flop field. [16] Most recently, a negative SMR was presented for another HM/AFM bilayer, $Pt/\alpha-Fe_2O_3$. [17], [18] Importantly, the presence of small net magnetization in $\alpha-Fe_2O_3$ layers enabled the magnetotransport properties of $\alpha-Fe_2O_3$ in a relatively small magnetic field to be revealed. Fisher *et al.* [17] noted sinusoidal oscillations of the Pt resistivity together with rotation of the Néel vector of the $\alpha-Fe_2O_3$ layer, with the amplitude of the SMR signal three times higher than that observed for the Pt/NiO system. In parallel, Cheng *et al.* demonstrated electrical switching of tristate Néel order in $Pt/\alpha-Fe_2O_3$ bilayers by measuring the Hall resistance. [18] Reversible and repeatable switching of the spin structure induced by electric current pulses was confirmed by direct observation of the domain structure with the use of a photoemission electron microscope. [19] These results, together with a recent theoretical and experimental demonstration of spin-pumping signals, inverse spin-Hall voltages [20], [21] and long-distance spin transport in hematite, [22], [23] indicate that the $Pt/\alpha-Fe_2O_3$ system has strong potential for antiferromagnetic spintronic applications.

In previous experimental studies concerning the magnetotransport properties of $Pt/\alpha-Fe_2O_3$, hematite layers with thicknesses of a few tens of nanometers were grown by sputtering [18], [19], [24] or pulsed laser deposition [17] on $Al_2O_3(0001)$ single crystals, followed by *in situ* room temperature deposition of thin Pt films. Whereas room temperature deposition of the Pt layer prevents reduction of the AFM hematite layer to ferrimagnetic magnetite ($Fe_3O_4$), [25], [26] the nonannealed Pt layer is structurally unstable, which can affect the magnetotransport properties of the Pt/α-Fe$_2$O$_3$ bilayer. In particular, Cheng *et al.* [18] reported current-induced triangular-shaped Hall resistance changes in the as-grown Pt/α-Fe$_2$O$_3$ bilayer grown by sputtering. They showed that triangular switching is an artifact that can be attributed to the current-driven migration of grain boundaries in the Pt layer and is not related to AFM switching, which confirms previous theoretical and experimental studies. [27] Interestingly, annealing of the sample by high current pulses was shown to result in single-pulse saturated, steplike switching of the AFM and improve the Pt stability. Notably, the postdeposition treatment temperature can influence the Pt/α-Fe$_2$O$_3$ interface, which is crucial for magnetotransport effects. Cheng *et al.* showed that interdiffusion of Fe and Pt at the Pt/α-Fe$_2$O$_3$ interface can lead to the formation of FePt alloy and induction of magnetization in Pt, which affects the magnetotransport properties of the system at low temperature. [24]

To dispel doubts about the stability of the Pt/α-Fe$_2$O$_3$ interface and its changes induced by annealing, a reversed α-Fe$_2$O$_3$/Pt structure can be considered. The use of the Pt layer as the substrate enables its deposition at elevated temperature and/or postdeposition annealing prior to the growth of α-Fe$_2$O$_3$, which provides the formation of a stable α-Fe$_2$O$_3$/Pt interface.

In our studies, we investigated the chemical and magnetotransport properties in the epitaxial α-Fe$_2$O$_3$/Pt structure, where hematite α-Fe$_2$O$_3$(0001) thin films with thicknesses of (6±0.3) nm and (15±0.5) nm were grown on a 7-nm thick epitaxial Pt(111) layer on MgO(111). Whereas surface-sensitive methods indicated only the hematite phase, Mössbauer spectroscopy measurements showed that the bulk-like hematite phase contributed 96% and 66% of the total spectral intensity for the thicker and thinner oxide layers, respectively. Although the thinner film contains almost a one-third contribution from a spinel phase, for simplicity, we refer to the thinner and thicker samples as α-Fe$_2$O$_3$(6)/Pt and α-Fe$_2$O$_3$(15)/Pt, respectively. An SMR study revealed that the chemical structure determines the magnetotransport properties of the α-Fe$_2$O$_3$/Pt bilayer. We noted a sign change of the SMR from positive to negative when the thickness of the oxide was increased, which was discussed in terms of the residual ferrimagnetic phase present in the thinner layer. Finally, for α-Fe$_2$O$_3$(15)/Pt, we demonstrated electrical switching. For the as-grown sample, we registered steplike, nondecaying switching between three antiferromagnetic order states, which demonstrates the stability of the interface in our samples.

## II. STRUCTURAL AND MAGNETIC CHARACTERIZATION

Hematite (α-Fe$_2$O$_3$), which is the most stable phase of iron oxide, crystallizes in the trigonal corundum structure with a hexagonal unit cell. Above the Morin transition temperature ($T_M$=250 K) [28] and below the Néel temperature ($T_N$=953 K), [28] the spins of the Fe$^{3+}$ ions are ferromagnetically aligned within the (0001) plane, whereas neighboring (0001) planes form "ABBA" stacking along the [0001] direction (*c* axis).

For $T_M < T < T_N$, the Dzyaloshinskii-Moriya interaction (DMI) induces slight canting of two AFM sublattices, which leads to the appearance of a weak ferromagnetism, with the ferromagnetic component almost perpendicular to one of the three easy axes, i.e., $[10\bar{1}0]$, $[0\bar{1}10]$ and $[\bar{1}100]$ within the (0001) plane. [29] Below $T_M$, hematite undergoes a transition to an easy-axis AFM with spins antiferromagnetically aligned along the c axis. [30] In thin films, with the decrease in hematite thickness, a decrease in $T_M$ or complete disappearance of the transition was observed. The transition was also shown to be restorable by doping. [31], [32], [33]

In our study, α-Fe$_2$O$_3$/Pt bilayers were prepared in a multichamber UHV system. A one-side polished MgO(111) single crystal was used as the substrate. Prior to deposition, the substrate was degassed for 30 minutes at 870 K. A 7 nm-thick epitaxial Pt(111) layer was deposited at a rate of 0.06 nm/min at room temperature and annealed at 800 K for 30 minutes to improve its surface quality. Hematite layers were grown via oxidation of magnetite Fe$_3$O$_4$(111) films. [27] First, magnetite layers were grown by reactive deposition of Fe in a molecular oxygen atmosphere under a partial pressure of 8x10$^{-6}$ mbar and a substrate temperature of 523 K. The samples were prepared with the use of the $^{57}$Fe isotope to enable Mössbauer spectroscopy studies. After deposition of 3 nm of metallic Fe (which corresponds to 6.3 nm of Fe$_3$O$_4$), half of the sample was protected by a shutter, while the deposition of Fe was continued on the uncovered part of the sample up to 7 nm (which corresponds to 14.7 nm of Fe$_3$O$_4$). As a result, two regions with different magnetite thicknesses were formed on the sample. Following Fe$_3$O$_4$ deposition, the sample was annealed at 770 K for 30 min. Oxidation of magnetite to hematite was performed by exposition of the Fe$_3$O$_4$ layers to oxygen, up to a total dose of 6075 L (2700 s at an oxygen partial pressure of 3x10$^{-5}$ mbar) while the substrate was kept at 770 K. The corresponding nominal thickness $d_h$ of the hematite films was 6.4 nm and 14.9 nm, with a thickness uncertainty of approximately ±5% (the accuracy of the deposition rate determination based on a quartz crystal monitor), and for simplicity, hereafter, the hematite thicknesses are referred to as 6 nm and 15 nm, respectively. After each preparation step, the films were characterized *in situ* by low energy electron diffraction (LEED). Additionally, at selected growth stages, analogous samples were characterized by scanning tunneling microscopy (STM) (see Supplemental Material). LEED patterns collected for the Pt, Fe$_3$O$_4$ and α-Fe$_2$O$_3$ surfaces revealed epitaxial growth of the layers and confirmed the formation of an α-Fe$_2$O$_3$/Pt bilayer with the epitaxial relation α-Fe$_2$O$_3$[$2\bar{1}\bar{1}0$]||Pt[$10\bar{1}$]||MgO[$10\bar{1}$] (Supplemental Material, Figure S1). From the surface-sensitive methods LEED and STM, no phases other than hematite could be concluded. Figure 1 shows 2θ/ω X-ray diffraction (XRD) scans of α-Fe$_2$O$_3$/Pt/MgO(111) for hematite thicknesses of 6 nm (blue) and 15 nm (red). For both parts of the sample, we observed reflections from the MgO substrate and interference of reflections from α-Fe$_2$O$_3$(0001) and Pt(111) layers. Interestingly, for the thinner hematite film, we noted an additional peak at a 2θ of approximately 57°, which can be a feature of the residual spinel structure in the oxide layer. [34] To unambiguously resolve the composition of the iron oxide films, we used conversion electron Mössbauer spectroscopy (CEMS). The undisputed advantages of MS, such as sensitivity to different chemical and structural atomic positions, make the method a perfect tool for characterization of iron oxides. [35] [26] The Mössbauer spectra of iron oxides with a long range magnetic order are magnetically split and possess a characteristic set of hyperfine parameters, i.e., isomer shift (IS), quadrupole splitting (QS) and hyperfine magnetic field ($B_{HF}$). Thus, numerical analysis of the spectrum allows identification and quantification of iron oxides. Particularly, a distinct feature of hematite is a pronounced quadrupole interaction that is present in compounds with trigonal symmetry, such as hematite, and almost vanishes for spinel cubic structures, such as magnetite or maghemite.

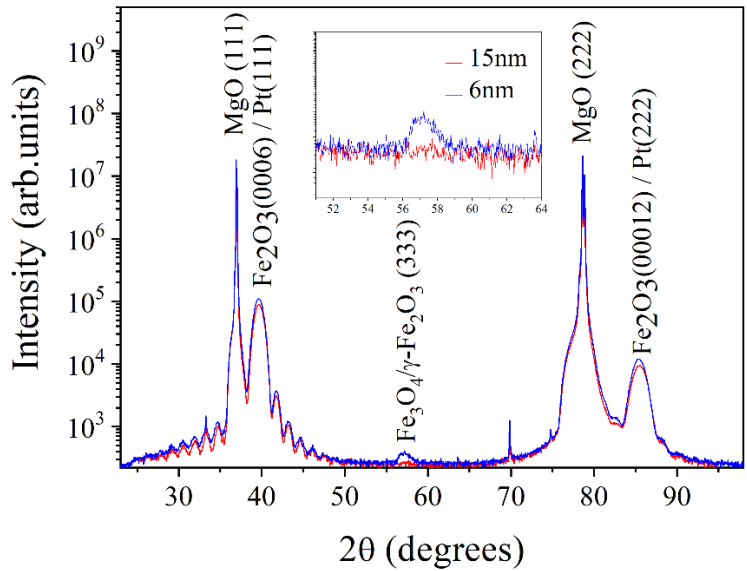

*Figure 1 XRD patterns of α-Fe$_2$O$_3$/Pt/MgO(111) with hematite layer thicknesses of 6 nm (blue) and 15 nm (red).*

Figures 2(a) and (b) show room-temperature CEMS spectra collected for the 6 nm and 15 nm films, respectively (for measurement details and fitting parameters, see Supplemental Materials, Table I). Both spectra are dominated by six line components (magenta shaded regions) with QS = -0.11 mm/s, IS = 0.37 mm/s and an average hyperfine magnetic field $B_{HF}$ of 50.9 T and 51.7 T for the thinner and thicker layers, respectively, which unambiguously identifies regular $Fe^{3+}$ sites in the corundum structure of hematite. [26] These components possess a bimodal $B_{HF}$ distribution, where the low intensity broader subcomponents with lower $B_{HF}$ (darker shaded regions in Fig. 2) represent surface hematite layers. [25] Additionally, the more complex spectrum for the 6 nm film exhibits a 30% contribution of sites (green shaded component in Fig. 2(a)) whose hyperfine parameters, in particular IS = 0.26 mm/s and QS = 0.03 mm/sec, are close to those of a spinel phase, namely, maghemite (γ-Fe$_2$O$_3$). The spinel phase (maghemite or nonstoichiometric magnetite, whose Mössbauer sextets are indistinguishable [36]) is a natural residue of the precursor phase for hematite that is obtained via oxidation of magnetite. The epitaxially stabilized spinel phase is more likely to occur in the thinner 6 nm layer than in the 15 nm film, as the stabilization of the spinel/hematite bilayer for the latter case would incur a much higher strain energy. A similar tendency was also found during direct growth of iron oxide films on Pt(111), where the early stages of growth favor maghemite and the thermodynamically stable phase, hematite, is stabilized in the thicker layer regime. [26] The broad distribution of $B_{HF}$ for this component is an interfacial effect previously reported in systematic studies of the oxidation of magnetite to hematite. [26] From line intensities in the spectra in Fig. 2 that were measured with the gamma-rays along the film normal it is clear that the AFM spins lay in the (0001) plane, as expected for hematite above the Morin temperature. Additional measurements at the grazing incidence as a function of the azimuthal angle indicate the three-fold symmetry of the in-plane spin distribution corresponding to the three easy axes (see Supplemental Materials for details).

Notably, down to the temperature of 110 K, we did not observe the Morin transition for either film (which, in the bulk, occurs at a temperature of approximately 250 K). The absence of the Morin transition in thin hematite films was previously reported for $\alpha$-$Fe_2O_3$ grown on Pt(111) [26] and $Al_2O_3$(001) [37] substrates. To summarize, whereas the 15 nm film is fully composed of antiferromagnetic hematite, in the 6 nm film, a 33% contribution of a ferrimagnetic spinel phase must be considered. Comparison of the volume-sensitive methods (XRD and CEMS) with LEED indicates that the spinel phase is localized in deeper layers. The presence of this non-hematite phase can be substantial for the magnetotransport properties of the $\alpha$-$Fe_2O_3$/Pt bilayer, as discussed below.

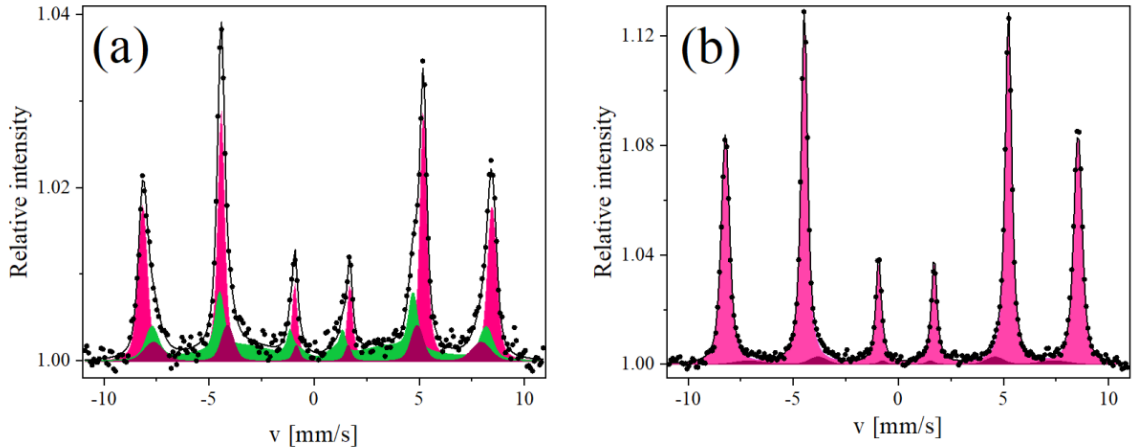

*Figure 2 Experimental CEMS spectra (points) of α-Fe$_2$O$_3$/Pt, and results of the best fits (lines) and their deconvolution into spectral components (shaded) for the 6 nm (a) and 15 nm (b) films. For the color code, see the text.*

## III. SMR MEASUREMENTS

For the magnetotransport experiments, we patterned Hall bar mesa structures on our sample with the use of photolithography and ion beam milling. The width and length of the Hall bars were 30 μm and 80 μm, respectively. Room-temperature magnetoresistance measurements were performed by rotating the external magnetic field **H**$_α$ relative to the [$\bar{1}$100] axis of α-Fe$_2$O$_3$ within the (0001) plane (Fig. 3a). The SMR measurements were performed as a function of magnetic field strength. Longitudinal resistances ($R_{xx}$) were determined from 4-wire resistance measurements in which a dc current density of 9.2 x 10$^{10}$ A/m$^2$ ($J_c$) was applied along the [$\bar{1}$100] crystal axis of α-Fe$_2$O$_3$ (x axis) (see Supplemental Materials for details of the measurements). Figure 3 shows the $R_{xx}(α)$ dependence obtained for the α-Fe$_2$O$_3$/Pt bilayers with hematite thicknesses of 15 nm (Fig. 3(b)) and 6 nm (Fig. 3(c)). For both hematite thicknesses, we noted SMR oscillations with a period of 180°. Interestingly, we observed a phase shift of 90° in the $R_{xx}(α)$ dependences measured for the $d_h$ of the 6 nm and 15 nm bilayers. Whereas for the thinner film, we noted $R_{xx}$ maxima for **H**$_α$ parallel/antiparallel to the x direction and longitudinal resistance minima for **H**$_α$ perpendicular to x (parallel/antiparallel to the y direction), the $R_{xx}$ maxima and minima for $d_h$ = 15 nm appear for **H**$_α$ ∥ ±y and **H**$_α$ ∥ ±x, respectively. The $R_{xx}(α)$ characteristic observed for the thicker hematite layer (Fig. 3(b)) results from the perpendicular alignment of the Néel vector with respect to the external magnetic field and is a fingerprint of AFM SMR (so-called "negative SMR"). [16] The amplitude of the AFM SMR obtained in our study for α-Fe$_2$O$_3$(15)/Pt(7) saturates at magnetic field of 2T (Fig. S3, Supplemental Materials), similarly to the previous studies obtained for reversed Pt/α-Fe$_2$O$_3$ structure [17]. At 2T the amplitude of SMR is approximately 1x10$^{-4}$. This value is higher than the SMR amplitude noted for the Pt(3.5)/NiO(120) bilayer registered at magnetic field of 15T. [16] For our thinner film, we noted a 90° phase shift in the SMR signal (Fig. 3(c)), i.e., the SMR exhibits the cos$^2$α dependence expected for FM/HM bilayers (positive SMR), in which the FM magnetization aligns parallel to **H**$_α$. [2] A positive SMR has been recently reported for AFMs, and its origin was attributed to the presence of either a ferromagnetic component at the AFM/HM interface or a canted AFM spin structure. [13], [38], [39], [40], [41] For bulk hematite, the small net magnetization induced by the DMI was shown to not contribute to the electrical response, which is dominated by the AFM Néel vector. [42] Thus, a negative SMR is expected for α-Fe$_2$O$_3$/Pt bilayers, as we found for the thicker hematite layer. In our study, the appearance of a positive SMR for the α-Fe$_2$O$_3$(6)/Pt bilayer is associated with a notable contribution of a spinel-like component detected in the Mössbauer spectrum, which we localize at the Pt/iron oxide interface. Both Fe$_3$O$_4$ and γ-Fe$_2$O$_3$ are ferrimagnets and can contribute to the positive SMR. The amplitude of the positive SMR obtained in our study for α-Fe$_2$O$_3$(6)/Pt is three times greater than that obtained for Pt(7)/Fe$_3$O$_4$(20)/MgO(001) [5] and twice as large as that obtained for the γ-Fe$_2$O$_3$/Pt bilayer. [43]

The SMR signal appears in thinner α-Fe$_2$O$_3$ sample for much weaker magnetic fields, i.e. we noted a clear oscillations of resistance for α-Fe$_2$O$_3$(6) at 0.05T while at the same magnetic field no SMR was noted for α-

$Fe_2O_3(15)$ (compare orange in Figure 3(b) and (c)). This confirms higher sensitivity of ferrimagnetic moment of $\alpha$-$Fe_2O_3(6)$ to the external magnetic field. Amplitudes of SMR at considerable magnetic fields (see Fig. S3 in Supplemental Materials) are comparable. This suggests that the effectiveness of the spin transfer from Pt layer to the $\alpha$-$Fe_2O_3(6)$ and $\alpha$-$Fe_2O_3(15)$ layer is similar. This result may not be surprising if we consider that both materials consist of $Fe^{3+}$ insulating oxide.

To exclude proximity-induced anisotropic magnetoresistance (AMR) contribution to the magnetoresistance in our samples we performed additional magnetoresistance measurements in which we rotated the external magnetic field within plane perpendicular to y direction (xz plane). [5] For both $\alpha$-$Fe_2O_3(6)$/Pt and $\alpha$-$Fe_2O_3(15)$/Pt heterostructures we did not observe oscillation of magnetoresistance within xz plane characteristic for AMR (Fig. S4 in Supplemental Materials) [5]. Thus, proximity-induced AMR do not contribute to the magnetoresistance in our study.

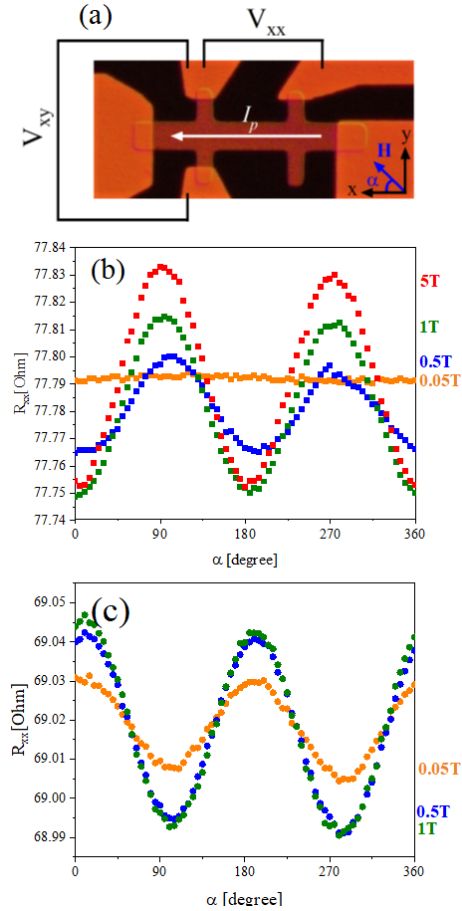

*Figure 3 (a) Schematic of the Hall bar structure prepared for $\alpha$-$Fe_2O_3$/Pt bilayers. During the SMR measurements, the external magnetic field was rotated in the $\alpha$-$Fe_2O_3$ (0001) plane (xy plane) relative to the [$\bar{1}$100] axis of $\alpha$-$Fe_2O_3$ (x axis). (b), (c) Longitudinal SMR respectively for the $\alpha$-$Fe_2O_3(15)$/Pt(7) and $\alpha$-$Fe_2O_3(6)$/Pt(7) performed for different strength of magnetic field.*

## IV. CURRENT-INDUCED SWITCHING OF THE NEEL ORDER IN α-Fe$_2$O$_3$(14.7)/Pt(7 nm)

To demonstrate the possibility of current-induced switching between the three easy axes of an AFM α-Fe$_2$O$_3$ layer, we patterned the α-Fe$_2$O$_3$(15)/Pt(7) bilayer into 8 terminal Hall crosses (Fig. 4(a)) in addition to the Hall bar described in the previous section. Three (10 μm wide) of four electrical paths were aligned along the easy axes of the α-Fe$_2$O$_3$ layer, i.e., along [10$\bar{1}$0], [$\bar{1}$100] and [0$\bar{1}$10] (black, green and red arrows in Fig. 4(a), respectively), whereas the fourth path (5 μm wide) was used to measure the Hall voltage (horizontal path, *y* direction in Fig. 4(a)). During the measurements, we used the following switching protocol. A set of twenty 1-ms-long current pulses (I$_p$) was applied along one of the easy hematite directions. After a 1-s delay, a reading current of 0.18 mA was applied along the [$\bar{1}$100] axis, and the transversal Hall voltage (V$_H$) was recorded. This series was repeated 15 times. Then, the series of current pulses were applied along the other easy axes. Figure 4(b) presents the change in the transverse resistance (ΔR$_{xy}$) as a function of the number of current pulses for different current densities *j*. Similar to a previous study performed for the reversed Pt/α-Fe$_2$O$_3$ structure, [18] we observed a tristate Hall resistance after the application of the series of current pulses along the three easy axes of hematite. According to previous theoretical and experimental studies, a pulse current applied along one of the easy axes of an AFM exerts a spin-orbit torque (SOT) on the Néel order parameter and aligns it parallel to I$_p$. [18] Application of current pulses along the [10$\bar{1}$0], [$\bar{1}$100] and [0$\bar{1}$10] axes results in stabilization of the high, intermediate and low R$_{xy}$ states, respectively. To confirm the direction of the Néel vector (**L**) corresponding to a given resistance state, we determined the transverse Hall resistance (R$_{xy}$) dependence on the α angle at an in-plane magnetic field of 0.8 T (Fig. 3(a)). Due to the spin-flop, at α = -30°, the Néel vector is aligned along the [0$\bar{1}$10] axis, whereas for α = 30° and α = 90°, the spins of hematite are parallel and antiparallel to [10$\bar{1}$0] and [$\bar{1}$100], respectively. The dependence of R$_{xy}$ on α confirms that the state with the highest resistance is obtained for **L**∥[10$\bar{1}$0] (Fig. 4(c), empty star), whereas the intermediate and lowest transverse resistances are generated for **L**∥[$\bar{1}$100] (Fig. 4(c), green triangle) and **L**∥[0$\bar{1}$10] (Fig. 4(c), red square), respectively. In agreement with the theory of a damping-like SOT, [44] together with an increase in the current density, we noted an increase in ΔR$_{xy}$ from 19 mΩ to 26 mΩ when j changed from 9.37 x 10$^{11}$ A/m$^2$ to 9.45 x 10$^{11}$ A/m$^2$. The current densities required for reorientation of the Néel vector are similar to the values reported for magnetization switching in FM/HM multilayers. [45], [46] Importantly, in our experiment, we noted rectangular, nondecaying switching characteristics with no traces of sawtooth features for the as-grown sample. This result is distinct from switching experiments performed for the reversed Pt/α-Fe$_2$O$_3$ structure, for which sawtooth-shaped switching of ΔR$_{xy$ was observed for the as-prepared sample and step-like switching appeared only after annealing of the sample with high current pulses. [18] We attribute this difference to the stable α-$Fe_2O_3$/Pt interface formed during the growth of our sample, which remained unaffected by the current pulse treatment.

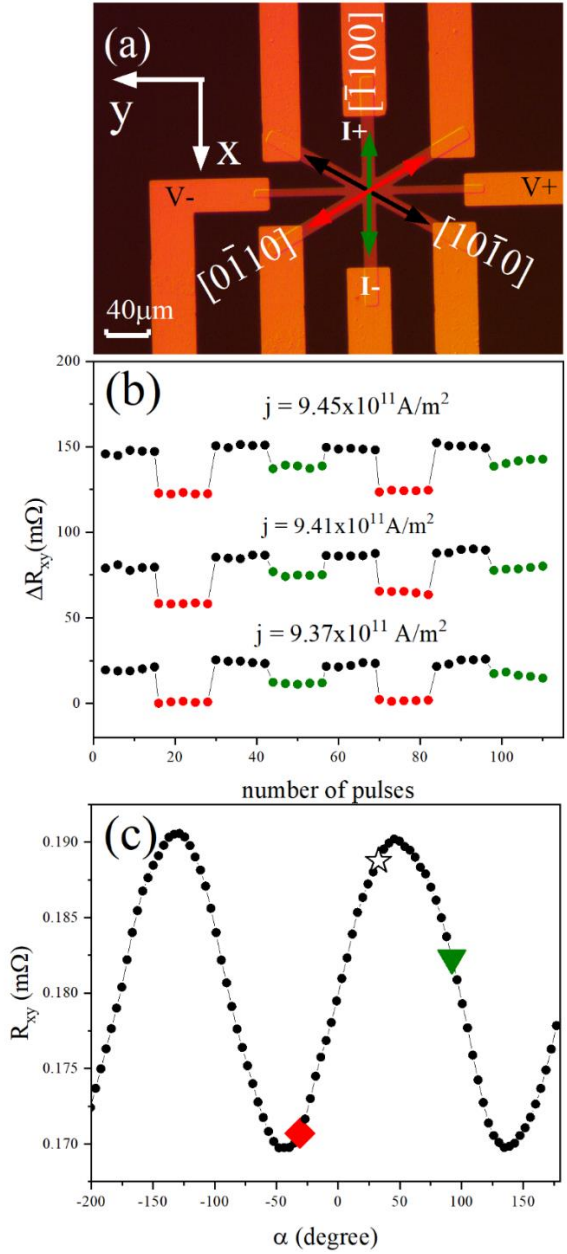

*Figure 4 (a) Optical microscopy image of the eight-terminal device used for the Néel order switching experiment on the α-$Fe_2O_3$(15 nm)/Pt bilayers. (b) Change in the transverse resistance as a function of the number of current pulses applied along the three easy axes of hematite. Measurements with different j are artificially offset for clarity. (c) Dependence of the transversal resistivity ($R_{xy}$) on angle α at $H_α$=0.8 T.*

## V. SUMMARY AND CONCLUSIONS

We investigated the structure and magnetotransport properties of α-$Fe_2O_3$/Pt bilayers with two iron oxide thicknesses, 6 nm and 15 nm. Mössbauer spectroscopy measurements showed that the hematite phase dominates for both oxide thicknesses; however, for the thinner film, interfacial contributions of a residual spinel structure are notable. The structural difference affects the magnetotransport properties of the α-$Fe_2O_3$/Pt bilayers. Whereas for the thicker film, we observed a negative SMR, which is characteristic of AFM/HM bilayers, the sign of the SMR for the bilayer with the thinner oxide film was positive. We attributed the change in the SMR sign to the interfacial ferrimagnetic order within the spinel phase.

Finally, for α-$Fe_2O_3$(15)/Pt, we demonstrated room-temperature switching of the Néel vector. The steplike nondecaying switching observed for the as-grown sample reveals the formation of a stable AFM/HM interface in our samples. We also note that such structures, with hematite grown on Pt, open the way to extending magnetotransport studies to more complex systems with double asymmetric HM/hematite/Pt interfaces.

**Acknowledgments**

This work was supported by Grant No. 2016/21/B/ST3/00861 funded by the National Science Centre, Poland, and Grant No. 2020/38/E/ST3/ 00086 funded by the National Science Centre, Poland. The authors acknowledge Felix Casanova and Luis Hueso for their technical expertise in the magnetotransport measurement. W.S. acknowledges Bekker grant No. PPN/BEK/2020/1/00118/DEC/1 from the Polish National Agency for Academic Exchange.